\documentstyle[aps,epsf,twocolumn]{revtex}
\input epsf.sty
\begin{document}
\draft
\flushbottom
\twocolumn[
\hsize\textwidth\columnwidth\hsize\csname @twocolumnfalse\endcsname

\title{Non-equilibrium effects in transport through quantum dots}
\author{E. Bascones$^1$, C. P. Herrero$^1$,
F. Guinea$^1$ and Gerd Sch\"on$^{2,3}$}
\address{
$^1$Instituto de Ciencia de Materiales, Consejo Superior de Investigaciones
Cient{\'\i}ficas,
Cantoblanco, E-28049, Madrid, Spain \\
$^2$Institut f\"ur Theoretische Festk\"orperphysik, Universit\"at
Karlsruhe, 76128 Karlsruhe, Germany\\
$^3$ Forschungszentrum Karlsruhe, Institut f\"ur Nanotechnologie,
76021 Karlsruhe, Germany}
\date{\today}
\maketitle
\tightenlines
\widetext
\advance\leftskip by 57pt
\advance\rightskip by 57pt

\begin{abstract}
The role of non-equilibrium effects in the conductance through 
quantum dots is investigated. Associated with
single-electron tunneling are shake-up processes and 
the formation of excitonic-like resonances. They change qualitatively the low
temperature properties of the system. We analyze by quantum Monte
Carlo methods the renormalization of the effective capacitance and the
gate-voltage dependent conductance. Experimental relevance is discussed.   
\end{abstract}

]
\narrowtext
\tightenlines
\section{Introduction.}
Quantum dots are paradigms to study the transition
from macroscopic to microscopic physics. At present, the role
of single-electron charging is well understood\cite{SET}.
Processes which otherwise are found in solids at the
single-atom level, such as
the Kondo effect, are being currently investigated\cite{Getal98}. 
Other atomic features,
like the existence of high spin ground states have also
been observed\cite{Oetal98}. The existence of many internal degrees of
freedom within the dot leads to a variety of effects reminiscent
of those found in large molecules. Shake-up processes, associated with
the rearrangement of many electronic levels upon the addition
of one electron, have been reported\cite{Setal97,Oetal99,Aetal97}.
%as well as the formation of bonding resonances in double well 
%systems\cite{Vetal95,Betal98,Oetal98b}.

In the present work, we will show that internal
excitations of the dot lead to non-equilibrium effects which
can substantially modify the transport properties.
In sufficiently small dots, the addition of a single electron may
cause significant charge rearrangements\cite{Setal96}, 
%Resonances associated to shake-up processes of the internal degrees
%of freedom of the dot were reported in\cite{Aetal97}. 
and the resulting change in the electrostatic potential  of the dot
modifies the electronic level structure. 
In the limit when the level separation is much smaller than other
relevant scales, this process leads to an   
^^ ^^ orthogonality catastrophe"\cite{A67}, first discussed in
relation with the sudden switching of a local potential in
a bulk metal. In addition, an electron tunneling event 
changing the charge of the dot is associated with
a charge depletion in the leads or in neighboring
dots. The attraction between the electron in the dot and the
induced positive charge leads to the formation of
an excitonic resonance, similar to the well known
excitonic effects in X-ray absorption\cite{ND69,M91}.
%A qualitative representation of these processes are shown in
%fig.[\ref{fig.sketch}]. 

The relevance of these effects for
tunneling processes in mesoscopic systems was first discussed
in refs.~\cite{UK90,UG91} (see also\cite{GBC98}). 
The existence of excitonic-like resonances
has indeed been reported in mesoscopic systems\cite{MVM97}.
The formation of bonding resonances has been observed 
in double well systems\cite{Vetal95}.   
Non-equilibrium effects like those studied here 
have also been discussed in Ref.~\cite{ML92} 
where transport through a tunnel junction 
via localized levels due to impurities was analyzed, and have been 
observed experimentally in Ref.~\cite{Getal94}.
% In this case charging effects do not need to be taken into account. 
% An excitonic resonance appears due to the Coulomb interaction 
% between the electron which tunnels to the localized level 
% and the conduction electrons in the leads.    
A different
way for modifying the conductance of quantum dots through the formation
of excitonic states has been proposed in\cite{PAK99}. In that case the
exciton is a real bound state, while the excitonic mechanism discussed
here is a dynamical process.                                
In some devices, the charge rearrangement may
take place far from the tunneling region. In this case 
no excitons are formed but the orthogonality catastrophe persists. 
This process has been discussed in relation to the measurement
of the charge in a quantum dot by the current through a
neighboring point contact\cite{Yetal95}.

We will study the simplest
deviations from the standard Coulomb blockade regime. Our analysis
is valid for quantum dots where the spacing between
electronic levels, $ \Delta \epsilon$, is much smaller than the
charging energy, $E_C$, or the temperature $T$. The non-equilibrium effects
discussed here will be observable if, in addition, the number of
electrons in the dot is not too large so that changes in the charge 
state lead to non-uniform redistributions of the charge. 
Thus, we will consider an intermediate situation between the
Kondo regime and orthodox Coulomb blockade (see below for estimates).
The processes that we consider will be present in double-dot
devices, but, for definiteness, we study here a single dot.

The main features discussed above can be described by a generalization of
the dissipative quantum rotor model\cite{UG91}, which has been
studied widely in connection with conventional Coulomb blockade
processes\cite{SZ90}. We present
a detailed numerical analysis of the generalized model,
along similar lines as previous work by some of us
on the dissipative quantum rotor\cite{HSZ99}.

The  paper is organized as follows:
In the next section, we show how to
estimate the parameters which characterize non-equilibrium effects.
The model is reviewed in section III, with emphasis on details
needed for the subsequent calculations. The numerical method is
presented in section IV.
In Secs. V and VI, we give results
for the renormalized capacitance and conductance.
In Sec. VII we discuss some possible experimental evidence of the effects studied here.We close with some conclusions.

\section{Non-equilibrium effects.}
\subsection{Inhomogeneous charge redistribution.}
The standard Coulomb blockade model assumes that, upon a
change in the charge state of the dot,
the electronic levels within 
a quantum dot are rigidly shifted by the charging energy, $E_C=e^2/2C$ with $C$ the capacitance of the dot . 
Deviations from this assumption
have been studied by means of an expansion in terms of $g^{-1}$, 
the inverse dimensionless
conductance, $g \sim k_{\rm F} l$, where $k_{\rm F}$ is the Fermi wave-vector,
and $l$ is the mean free path\cite{BMM97}. It is also assumed that
$k_{\rm F}$ is small compared to the inverse Thomas-Fermi screening length,
$k_{\rm TF} = \sqrt{4 \pi e^2 N ( \epsilon_{\rm F} )/ \epsilon_0}$.
To lowest order beyond standard Coulomb blockade effects, the change
in the charge state of the dot  
leads to an inhomogeneous potential, and induces a 
term in the Hamiltonian, which can be written as\cite{BMM97,G00}:
\begin{equation}
{\cal H}_{\rm int} = ( Q - Q_{\rm offset} )
\int \psi^{\dag} ( {\bf \vec{r}} ) U ( {\bf \vec{r}} ) 
\psi ( {\bf \vec{r}} ) d^d {\bf r} \; .
\label{potential}
\end{equation}
Here $Q_{\rm offset}$ denotes offset charges in the
environment, the operator $Q$ measures
the total electronic charge in the dot,
and $\psi^\dag ( \bf{\vec{r}} )$ creates an electron 
at position $\bf{\vec{r}}$.
The potential $U({\bf \vec{r}} )$  modifies 
the constant shift of the energy levels
of the dot assumed in the standard Coulomb blockade model. 
It appears due to the restricted geometry of the dot. 
After a charge tunnels into the dot, a pile-up of electrons
at the surface of the dot is induced. As a result there is a net attraction of
electrons towards the surface, besides a constant shift given by
$e^2 / 2C$. In general, the potential $U$ in eq.~(\ref{potential}) can be
obtained from the Hartree approximation (in the Thomas-Fermi limit)
for dots and leads of arbitrary shape. For a
spherical dot of radius $R$
the potential $U ( \bf{\vec{r}} )$ has the simple form\cite{BMM97}:     
\begin{equation}
U ( {\bf \vec{r}} ) = - \frac{e^2 e^{- k_{FT} ( R - r )}}
{\epsilon_0 k_{\rm TF} R^2} + K  
\end{equation}
and, for a two-dimensional circular dot:
\begin{equation}
U ( {\bf \vec{r}} ) = - \frac{e^2}{2 \epsilon_0 k_{\rm TF} R
\sqrt{R^2 - r^2}} + K \; .
\end{equation}                                                

K is a constant which ensures $\langle U({\bf \vec{r}})\rangle=0$.
Non-equilibrium effects arise because
the potential in eq.~(\ref{potential}) is time dependent,
as it changes upon the addition of electrons to the dot.
Hamiltonians with terms such as eq.~(\ref{potential})
were first discussed in\cite{UG91} (see next section). 
Note that the potential is
localized in the surface region, where the tunneling electron is
supposed to land. Finally, the potential is attractive, leading 
to the localization of the new electron near the surface, giving
rise to excitonic effects.

Other non-equilibrium effects can arise 
if the charge of the dot induces inhomogeneous potentials in other
metallic regions of the device. In this case, the only effect
expected is the orthogonality catastrophe (see below), due to the shake-up of
the electrons both in the dot and in the other regions. We take this
possibility into account in the analysis in the
following sections.

\subsection{Effective tunneling density of states.}

In the absence of non-equilibrium effects, 
the conductance of a junction between
the dot and the leads is  
\begin{equation}
g =  \frac{2 e^2}{h} \sum_{\rm channels} | t_i |^2 
N_{i,{\rm lead}} ( E_{\rm F} ) N_{i,{\rm dot}} ( E_{\rm F} ) \; ,
\label{Bardeen}
\end{equation}
where the summation is over the channels, 
$t_i$ is the hopping matrix element through channel $i$,
$N(E_{\rm F})$ is the density of states at the Fermi level,
and we use the standard theory of
tunneling in the weak transmission limit\cite{BMS83}.
Eq. (\ref{Bardeen}) implicitly assumes a constant density
of states, as appropriate for a metallic contact.

The non-equilibrium effects to be considered can be taken into account
through a modification of the effective tunneling density of
states\cite{UG91,DRG98}. In this case the electron propagators in
eq.~(\ref{Bardeen}) are the non-equilibrium ones, 
in an analogous way to the modifications
required in the study of X-ray absorption spectra of core levels in
metals\cite{ND69}, or tunneling between Luttinger 
liquids\cite{KF92,MG93,SK97}.
As in those problems, we can distinguish two cases:

i) The analogue of the X-ray absorption process: The charging of the dot
leads to an effective potential which modifies the electronic levels.
At the same time, an electronic state localized in a region within the
range of the 
potential is filled. The interaction between the electron
in this state and the induced
potential must be taken into account (the excitonic effects, in the
language of the Mahan-Nozi\`eres-de Dominicis theory).

ii) The analogue of X-ray photoemission: The charging process leads
to a potential which modifies the electronic levels. The tunneling
electron appears in a region outside the range of this potential.
Only the orthogonality effect caused by the potential needs to be
included.

Taking into account the distinctions between these two possibilities,
the effective (non-equilibrium) 
density of states in the lead and the dot becomes
(omitting the channel index, $i$):
%\begin{eqnarray}
%D_{\rm eff} ( \omega ) &= &
\begin{eqnarray}   
D_{\rm eff} ( \omega )& = &\int_0^\omega d \omega' N_{\rm dot}^{\rm empty} (
\omega' )N_{\rm lead}^{\rm occ} ( \omega - \omega' ) \nonumber \\ 
& \propto &  |\omega |^{1-\epsilon}   
\label{effdosnew}
\end{eqnarray}
with $\epsilon$ given by 
\begin{eqnarray}
\epsilon = \left\{ \begin{array}{ll}  
  \sum_{j=1,2}  2 \frac{\delta_j}{\pi} -  \left(
\frac{\delta_j}{\pi} \right)^2 &  
\rm{\left( excitonic \right.} \\ &\rm{\left. resonance \right)} \\
 -\sum_{j=1,2}  \left(
\frac{\delta_j}{\pi} \right)^2   &  
\rm{\left( orthogonality \right.}  \\ & \rm{\left.
catastrophe \right)} \end{array} \right.
\label{effdos1}
\end{eqnarray}    
%&\left\{ \begin{array}{cc}
%| \omega |^{1 + \sum_{j=1,2} - 2 \frac{\delta_j}{\pi} + \left(
%\frac{\delta_j}{\pi} \right)^2}  &
%\rm{\left( excitonic \right.} \\ &\rm{\left. resonance \right)} \\
%| \omega |^{1 + \sum_{j=1,2}  \left(
%\frac{\delta_j}{\pi} \right)^2}  
%&\rm{\left( orthogonality \right.}  \\ & \rm{\left.
%catastrophe \right)} \end{array} \right.
%\label{effdos1}
%\end{eqnarray}
Here  $\delta_j$ is the phase shift induced by the 
new electrostatic potential in the lead states ($j=1$) or in the
dot states ($j=2$).  
The exponent is positive, $\epsilon > 0$, if excitonic effects prevail, while
$\epsilon < 0$ if the leading process is the orthogonality catastrophe.
%It is straightforward to extend this analysis to a double dot system.

We can get an accurate estimate for $\epsilon$ in the simple cases
of a spherical or circular quantum dot decoupled from other metallic
regions discussed in\cite{BMM97}. We assume that tunneling
takes place through a single channel, and the contact is
of linear dimensions $\propto k_{\rm F}^{-1}$.
In Born approximation the effective phase shift becomes:
\begin{equation}
\delta \approx N ( \epsilon_{\rm F} ) \int_{\Omega} U ( {\bf \vec{r}} )
d^d {\bf r} \approx \left\{ \begin{array}{ll}
\frac{e^2 N ( \epsilon_{\rm F} )}{\epsilon_0 k_{FT} k_{\rm F}^3 R^2}
&\rm{spherical \, \, dot} \\ 
\frac{e^2 N ( \epsilon_{\rm F} )}{\epsilon_0 k_{FT} k_{\rm F} R} 
&\rm{circular \, \, dot} \\ \end{array} \right.
\end{equation} 
where $\Omega$ is the region where tunneling processes to the leads
are non negligible, typically
of dimensions comparable to $k_F^{-1}$.
Note that, for a very elongated dot (d=1), the phase shift will
not depend on its linear size. As mentioned earlier, the leads
can modify significantly these estimates. The tunneling electron can 
be attracted to the image potential that it induces, enhancing
the excitonic effects ($\epsilon > 0$). On the other hand, shake-up
processes in metallic regions decoupled from the tunneling processes 
will increase the orthogonality catastrophe, without contributing
to the formation of the excitonic resonance at the Fermi energy.

\section{The model.}
The shake-up processes mentioned in the preceding section
are described by the Hamiltonian\cite{UG91}:
\begin{eqnarray}
{\cal H} &= &{\cal H}_Q + {\cal H}_{\rm R} + {\cal H}_{\rm L} +
{\cal H}_{\rm T} + {\cal H}_{\rm int}  \nonumber \\
{\cal H}_Q &= &\frac{( Q - Q_{\rm offset} )^2}{2 C}  \nonumber \\
{\cal H}_i &= &\sum_k \epsilon_{k,i} c^\dag_{k,i} c_{k,i}
\;\; , \;\; i = {\rm L,R} \nonumber \\
{\cal H}_{\rm T} &= &t e^{i \phi} \sum_{k,k'} c^\dag_{k,{\rm R}}
c_{k',{\rm L}} + h. c.  
\nonumber \\ 
{\cal H}_{\rm int} &= &( Q - Q_{\rm offset} )
 \sum_{k,k'} \left( V^R_{k,k'} c^\dag_{k,{\rm R}} c_{k',{\rm R}}
- V^L_{k,k'} c^\dag_{k,{\rm L}} c_{k',{\rm L}} \right) \nonumber \\
\label{Hamiltonian}
\end{eqnarray}
 Here  $\left[ \phi , Q \right] = i e$. The Hamiltonian separates the junction degrees of freedom into
a collective mode, the charge $Q$, and the electron degrees of freedom
of the electrodes and the dot. This separation is standard
in analyzing electron liquids, where collective charge oscillations
(the plasmons) are treated separately from the low-energy electron-hole
excitations. In our case, this implies that only those states
with energies lower than the charging energy are to be included
in ${\cal H}_i$, ${\cal H}_T$ and ${\cal H}_{int}$
in eq.(\ref{Hamiltonian}). Higher electronic states
contribute to the dynamics of the charge, described by
${\cal H}_Q$. The Hamiltonian, eq.(\ref{Hamiltonian}), suffices
to describe transport processes at voltages and temperatures
smaller than the charging energy.

In the following, we will express the offset charge $Q_{\rm offset}$,
introduced in eq.~(\ref{potential}),
by the dimensionless parameter $n_e = Q_{offset} / e$.
By $V_{k,k'}$ we denote the 
matrix elements of $U ( {\bf \vec{r}} )$ in the basis of the
eigenfunctions near the Fermi level. We allowed that inhomogeneous
potentials can be generated on both sides of the junction. We assume that tunneling can take place through several channels (index channel has been omitted). The transmission through each channel should be small for perturbation theory to apply.

The electrical relaxation associated with the tunneling process takes place
in two  stages. In the first, the tunneling electron is screened by the 
excitation of plasmons, forming the screened Coulomb potential. The
time scale for this process is of the order of the inverse plasma
frequency. Next, the screened Coulomb  
potential excites electron-hole pairs. As the electrons at the Fermi level have a much 
longer response time, they feel this change as a sudden and local perturbation.   

We will restrict ourselves to  the regime where the level spacing is
small, $\Delta\epsilon \ll T,E_c$. 
Using standard techniques\cite{SZ90,AES82}, we can integrate out the
electron-hole pairs  
and describe the system in terms of the phase $\phi$ and charge $Q$ 
only. 
This procedure leads to retarded interactions which
are long-range in time, as the electron-hole pairs have a
continuous spectrum 
down to zero energy. It is best to describe the resulting model
within a path-integral formalism. Because of the non-equilibrium effects 
the effective action is
a generalization of that derived for tunnel junctions\cite{BMS83,AES82}
\begin{eqnarray}
{\cal S}[\phi] &=& \int_0^{\beta} d \tau \frac{1}{4 E_C}  
\left( \frac{\partial \phi}{\partial \tau} \right)^2 + \nonumber \\
& &\alpha \int_0^{\beta} d \tau \int_0^{\beta} d \tau' 
 E_C^{\epsilon} \left( \frac{\pi}{\beta} \right)^{2-\epsilon}
\frac { 1 - \cos [ \phi ( \tau ) - \phi ( \tau' ) ] }
{\sin^{2 - \epsilon} [ \pi ( \tau - \tau' ) / \beta ]}  \; . \nonumber \\
\label{action}
\end{eqnarray}
It describes the low energy processes below
an upper cutoff of order of the unscreened charging energy,
$E_C$.
The parameter 
$\alpha \propto t^2 N_R ( \epsilon_{\rm F} ) N_L ( \epsilon_{\rm F} )$
is a measure of the high temperature conductance, 
$g_0$, in units of $e^2 / h$:
\begin{equation}
\alpha = \frac{g_0}
{4 \pi^2 ( e^2 / h )} \; .
\end{equation}
Note that the definition of the action, eq.~(\ref{action}), does not
allow us to study temperatures much higher than $E_C$.
The kernel which describes the retarded
interaction is given by the effective
tunneling density of states, eq. (\ref{effdosnew}).
The value of $\epsilon$ is the anomalous exponent
in the tunneling density of states, given in eq. (\ref{effdosnew}).

The action (\ref{action}) has been studied  extensively for
$\epsilon = 0$\cite{SZ90,HSZ99,FSZ95,WEG97}, describing charging
effects in  the single-electron transistor in the usual limit where
the electrodes are assumed to be in equilibrium. If
$\epsilon > 0$, the model 
has a non-trivial phase transition\cite{K77,GS86,SG97}. In this case, for
$\alpha > \alpha_{crit} \sim 2 / ( \pi^2 \epsilon )$, the system 
develops long-range order when $T = 1/\beta \rightarrow 0$, leading to
phase coherence and a diverging conductance.

\section{Computational method}

For a given offset charge $n_e$, the grand partition 
function can be written in terms of the phase $\phi$, as
a path integral \cite{SZ90}:
\begin{equation}
  Z(n_e) = \sum_{m=-\infty}^{\infty} \exp(2 \pi i m n_e) \, 
	    \int {\cal D} \phi \, \exp \left( - S[\phi] \right)
  \hspace{.2cm} ,
  \label{part1}
\end{equation}
where $m$ is the winding number of $\phi$,
and the paths $\phi(\tau)$ satisfy in sector $m$ the
 boundary condition
$\phi(\beta) = \phi(0) + 2 \pi m$.

The effective action and partition function can be rewritten
 in terms of the phase fluctuations 
$  \theta(\tau) = \phi(\tau) - 2 \pi m \tau / \beta$,
 with boundary condition
$\theta(\beta) = \theta(0)$, in the form
\begin{equation}
Z = \sum_{m=-\infty}^{\infty}  
   \exp(2 \pi i m n_e) \,  I_m(\alpha,\epsilon,\beta) \hspace{.2cm} .
   \label{part2}
\end{equation}
The coefficients $I_m(\alpha,\epsilon,\beta) = \int {\cal D} \theta 
 \, \exp \left( - S_m[\theta ] \right) $
 are to be evaluated with the effective action
$ S_m[\theta(\tau)] = S[\theta(\tau) + 2 \pi m \tau / \beta] $. 
They depend on the winding number
$m$, the temperature, and the dimensionless parameters $\alpha$
and $\epsilon$, but are independent of the offset charge $n_e$.
This means that the problem reduces, from a computational point of view,
 to the calculation of the relative values of
$I_m(\alpha,\epsilon,\beta)$, which can be obtained from  
Monte Carlo (MC) simulations apart from an overall normalization
constant \cite{HSZ99,WEG97}.
The partition function is even and periodic with respect to $n_e$,
$Z(n_e) = Z(-n_e) = Z(n_e + 1)$, and therefore
one can restrict the analysis to the range $0 \le n_e \le 0.5$.

The MC simulations have been carried out by the usual
discretization of the quantum
paths into $N$ (Trotter number) imaginary-time slices \cite{S93}.
In order to keep roughly the same precision in the calculated
 quantities,  as the temperature is reduced, the number of time-slices
$N$ has to increase as $1 / T$.
 We have found that a value $N = 4 \beta E_C$ is sufficient to
 reach convergence of $I_m$.
 Therefore, the imaginary-time slice employed in the discretization of the
 paths is $\Delta \tau = \beta / N = 1 / (4 E_C) $. 

When discretizing the paths $\phi(\tau)$ into $N$ points,
 it is important to treat correctly the 
$| \tau - \tau' | \rightarrow 0$ divergence that appears in
 the tunneling term $S_t[\phi]$ of the effective action 
[second term on the r.h.s. of Eq.\,(\ref{action})].
This divergence can be handled as follows. 
 In the discretization procedure, the double integral
in $S_t[\phi]$ translates into a sum extended
to $N^2$ two-dimensional plaquettes, each one with area
$(\Delta \tau)^2$. The above-mentioned divergence appears
in the $N$ ``diagonal'' terms ($\tau = \tau'$) and can be dealt
with by approximating the integrand close to $\tau = \tau'$  by 
$ E_C^{\epsilon}  |\tau - \tau'|^{\epsilon} (d \phi / d \tau)^2 / 2$.
Thus, by integrating this expression over the ``diagonal'' plaquette
$(i,i)$, with $1 \le i \le N$, one finds that its contribution
to $S_t[\phi]$ is given by
\begin{equation}
      \Delta S_t (\tau_i,\tau_i) =  2 \, E_C^{\epsilon}
          \, \frac{\alpha}{\epsilon + 1}  
      \left(  \frac {\Delta \tau}{2} \right)^{\epsilon + 2}
       \left(  \frac {d \phi}{d \tau} \right)^2_{\tau = \tau_i}
  \hspace{.2cm} ,
  \label{aprox2}
\end{equation}   
which is regular for $\epsilon \neq -1$. The error introduced by
this replacement in the discretization procedure is of the same
order as that introduced by the usual discretization of the
``non-diagonal'' terms.  We have checked that the results
of our Monte Carlo simulations obtained by using this procedure 
converge with the Trotter number $N$.

The partition function in Eq.(\ref{part2}) has been sampled by
the classical Metropolis method \cite{BH88}
for temperatures down to $k_{\rm B} T = E_C / 200$.
A simulation run proceeds via successive MC steps.
In each step, all path-coordinates (imaginary-time slices) 
are updated.  For each set of parameters ($\alpha, \epsilon, T$),
the maximum distance allowed
for random moves was fixed in order to obtain an acceptance ratio
of about $50 \%$.  Then, we chose a starting configuration for 
the MC runs after system equilibration during about $3 \times10^4$ MC
steps. Finally, ensemble-averaged values for the quantities
of interest were calculated from samples of $\sim 1 \times10^5$
quantum paths.
More details on this kind of MC simulations can be found 
elsewhere \cite{HSZ99}.

\section{Renormalization of the capacitance.}

We will study the capacitance renormalization for tunneling conductance 
$\alpha > 0$  by calculating the effective charging energy
$ E_C^*(T) = e^2 / 2 C^*(T)$, which can be obtained
as a second derivative of the free energy
$F = - k_{\rm B} T \ln Z$:
\begin{equation}
   E_C^*(T) =  \frac{1}{2}  \left. 
       \frac{\partial^2 F}{\partial n_e^2} \right|_{n_e=0}
          \hspace{.2cm} .
   \label{ec1}
\end{equation}	
At high temperatures, the free energy $F(n_e)$ depends weakly on
$n_e$, and the curvature [i.e., $E_C^*(T)$] approaches zero.
At low $T$, and for weak tunneling ($\alpha \ll 1$), it coincides
with the usual charging energy $E_C$.
By using Eqs.\,(\ref{part1}) and (\ref{ec1}) 
this renormalized charging energy can readily be expressed as
\begin{equation}
E_C^*(T) =
2 \pi^2 k_{\rm B} T \langle m^2 \rangle_{n_e=0} \;,
   \label{ec2}
\end{equation}
where $\langle m^2 \rangle_{n_e=0}$ is the second  moment of
 the coefficients $I_m$.

The correlation function in imaginary time  $G(\tau)$, that
will be used below to calculate the conductance, can be
calculated from the MC simulations as
$G(\tau) = \langle \cos[\phi(\tau) - \phi(0)] \rangle$.
This means in our context:
\begin{figure}
\epsfxsize=\hsize
\centerline{\epsfbox{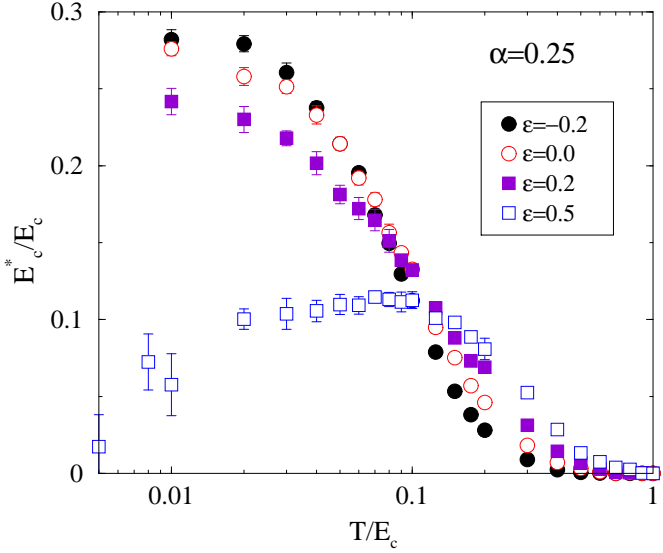}} %\vspace{-1.5cm}
\caption{Effective charging energy of the single
electron transistor for different values of $\epsilon$ and
$\alpha = 0.25$, as a function of temperature.}
\label{fig:eca25}
\end{figure}                                                        
\begin{eqnarray}
  G(\tau) & = & \frac{1}{Z}  \sum_{m=-\infty}^{\infty} \exp(2 \pi i m n_e)  
           \, \times    \nonumber  \\ 
       & & \int {\cal D} \phi \, e^{- S[\phi]} \cos[\phi(\tau) - \phi(0)]
  \hspace{.2cm} .
  \label{gtau}
\end{eqnarray}

A number of features, directly related to the free energy
of the model given by action (\ref{action}), are reasonably well
understood for $\epsilon = 0$. The effective charge induced by an
 arbitrary offset charge, $n_e$, has been extensively analyzed.
 A number of analytical schemes
give consistent results in the weak coupling ($\alpha \ll 1$) 
regime\cite{GG98,KS98}. These calculations have been extended
to the strong coupling limit by numerical methods\cite{HSZ99}.

\begin{figure}
\epsfxsize=\hsize
\centerline{\epsfbox{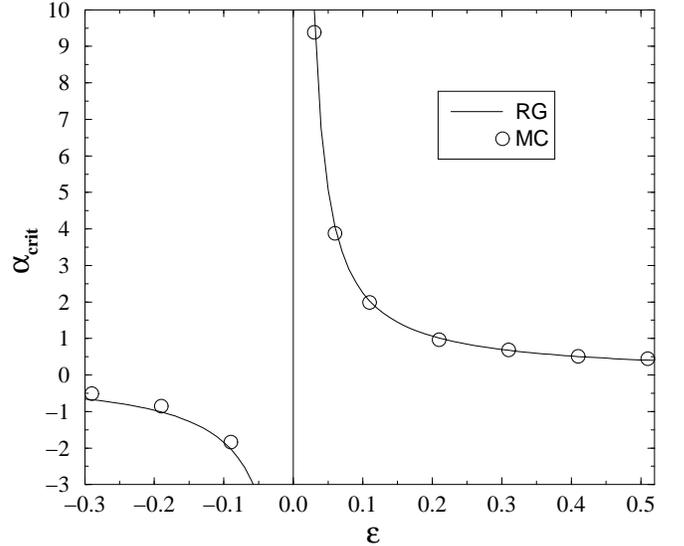}} %\vspace{-1.5cm}
\caption{Critical line determined by the RG calculation: $\alpha_{crit}=2/(\pi^{2}\epsilon)$ and values calculated fitting
the numerical data for $E^*_C ( \alpha , \epsilon )$ to the 
expression (\ref{alphacrit}).}
\label{fig:alphacrit}
\end{figure}

In fig.~(\ref{fig:eca25}), we present results for the effective
charging energy  $E^*_C$
as a function of temperature, and for different values of $\epsilon$.
The value of $E^*_C$ is enhanced for $\epsilon < 0$, where the orthogonality
catastrophe dominates the physics. A positive $\epsilon$ reduces the
effective charging energy, and, beyond some critical
value (see discussion below), $E^*_C$
scales towards zero as the temperature is decreased, showing
non-monotonic behavior.
The same trend can be appreciated in fig.~(\ref{fig:ect02}), where the
effective charging energy at low temperatures is plotted as a function
of $\alpha$.                                                 
Renormalization group arguments\cite{K77,GS86} show that
$E_C^* ( \alpha , \epsilon )$ 
should go to zero for $\alpha > \alpha_{crit} ( \epsilon )$.
We have checked the consistency of this prediction with
the numerical results by fitting
the values of $E^*_C ( \alpha )$ at low temperatures by the expression
expected from the scaling analysis near the transition\cite{K77}:
\begin{figure}
\epsfxsize=\hsize
\centerline{\epsfbox{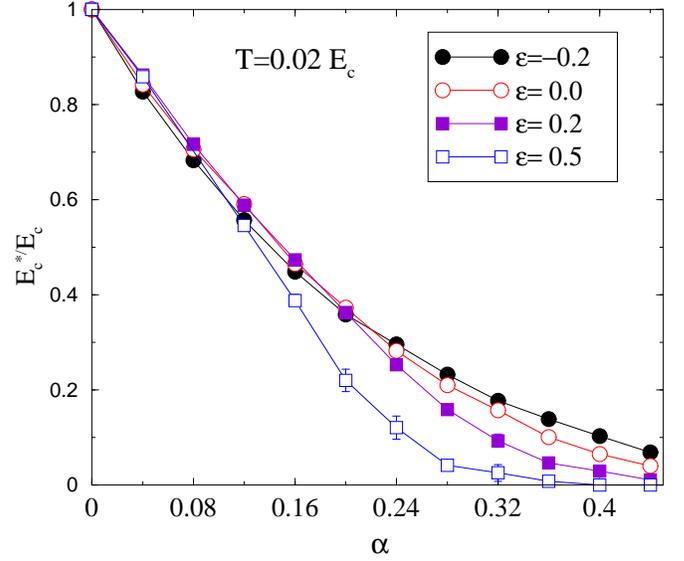}} %\vspace{-1.5cm}
\caption{Effective charging energy of the single
electron transistor for different values of $\epsilon$, as
a function of $\alpha$ for $T / E_C = 0.02$.}
\label{fig:ect02}
\end{figure}                                                      
\begin{equation}
E_C^*  ( \alpha , \epsilon ) 
= \left[ 1 - \frac{\alpha}{\alpha_{crit} ( \epsilon )} \right]^{
\frac{1}{\epsilon}}
\label{alphacrit}
\end{equation}
In this expression we use $\alpha_{crit} ( \epsilon )$ as the only adjustable
parameter.
The results are shown in fig.~(\ref{fig:alphacrit}).
Note that the same equation (\ref{alphacrit}) can be used to fit
the results for $\epsilon < 0$, if one uses a negative value
for $\alpha_{crit}$. There is no phase transition, however,
for $\epsilon < 0$. 
The results show that the present calculations are very accurate even for
relatively large values of $\alpha$, where $E^*_C$ converges at very
low temperatures.                                                     

\section{Evaluation of the conductance.}
\subsection{$\epsilon = 0$}
The conductance of the single-electron
transistor is notoriously more difficult to
calculate than standard thermodynamic
averages. It cannot be derived in a simple fashion from the partition
function, and requires the analytical continuation of the response
functions from imaginary to real times or frequencies\cite{GLT22}.
Hence, there are no comprehensive results valid for the whole
range of values of $\alpha, T/E_C$ and $n_e$. For $\epsilon = 0$, 
the conductance $g(T)$ can be written as\cite{ZS91,IZ92,G99}:
\begin{equation}
g ( T ) = 2 g_0 \beta \int_0^\infty \frac{d \omega}{2 \pi}
\frac{\omega S ( \omega )}{e^{\beta \omega} - 1}
\label{conductance}
\end{equation}
where $g_0$ is the normal state conductance,
and $S ( \omega )$ is related to the correlation function
in imaginary time\cite{W93}:
\begin{eqnarray}
G ( \tau ) &= 
&\langle e^{i \phi ( \tau )} e^{-i \phi ( 0 )} \rangle \nonumber \\ 
&= &\frac{1}{2 \pi} \int_0^\infty d \omega \,
      \frac{e^{(\beta - \tau ) \omega} + 
      e^{\tau \omega}}{e^{\beta \omega} - 1} A ( \omega )
\end{eqnarray}
and
\begin{equation}
A ( \omega ) = ( 1 - e^{- \beta \omega} ) S ( \omega )
\end{equation}
In the previous expressions, the charging energy is the natural cutoff
for the energy integrals.

At high temperatures, $\beta E_C \sim 1$, we can expand in eq.
(\ref{conductance}):
\begin{equation}
g ( T ) \approx 2 g_0 \beta \int_0^{\infty}  \frac{d \omega}{2 \pi}
\left[ \frac{1}{\beta} - \frac{\beta \omega^2}{24} + \frac{7 \beta^3
\omega^4}{5760} + ... \right] \frac{e^{(\beta \omega )/2} A ( \omega )}
{e^{\beta \omega} - 1}
\end{equation}
so that:
\begin{equation}
g ( T ) \approx g_0 \left[ G \left( \frac{\beta}{2} \right)
- \frac{\beta^2}{24}  G''  \left( \frac{\beta}{2} \right)
+ \frac{7 \beta^4}{5760} G^{iv}  \left( \frac{\beta}{2} \right)
+ ... \right]
\end{equation}

At low temperatures, $\beta E_C \gg 1$, the conductance is dominated
by the low energy behavior of $A ( \omega )$ or, alternatively,
$S ( \omega )$. To lowest order, we expect an expansion
of the form:
\begin{equation}
S ( \omega ) \approx  2 \pi \delta ( \omega - \omega_0 ) + A | \omega | + ...
\end{equation}
where $\omega_0$ is an energy of the order of the renormalized
charging energy (or zero at resonance, $n_e \approx 1 / 2$), and
$A$ is a constant which describes cotunneling processes\cite{AN90}.
Inserting this expression in eq.~(\ref{conductance}), we obtain:
\begin{equation}
g ( T ) \approx g_0 \frac{2\beta \omega_0}{ e^{\beta \omega_0} - 1} + g_0
\frac{A}{\pi \beta^2} \int_0^{\infty} dx \frac{x^2}{e^x -1} + ...
\label{glowt}
\end{equation}
while, on the other hand:
\begin{equation}
G \left( \frac{\beta}{2} \right) \approx  2e^{-\frac{1}{2} \beta
\omega_0} + \frac{A}{\pi} \left( \frac{2}{\beta} \right)^2 + ...
\label{Glowt}
\end{equation}
If $\omega_0 = 0$, both $g$ and $g_0 G ( \beta / 2 )$ have the
same limit, as $T \rightarrow 0$. When $\omega_0 \ne 0$ the leading
term goes as $T^2$ (cotunneling) in both cases, with prefactors 
equal to $2.404 A / \pi$ and $4 A / \pi$, respectively.

 From the above discussion of the relation between the high- and 
low-temperature behavior of $g ( T )$ and $G ( \beta / 2 )$, we find
that the interpolation formula
\begin{equation}
g ( T ) \approx g_0 G \left( \frac{\beta}{2} \right)
\label{inter}
\end{equation}
should give a reasonable approximation over the entire range
of parameters (note that the above discussion is independent
of the values of $\alpha$ and $n_e$). Eq. (\ref{inter})
is consistent with the the main
physical features expected both in the high and low temperature limits,
at and away from resonance. The advantage of using
$G ( \beta / 2 )$ is that it can be computed, to a high
degree of accuracy, by standard Monte Carlo techniques, as
it does not require to continue the results to real times.
A similar approximation, used to avoid inaccurate analytical
continuations has been applied for bulk systems in Ref.~\cite{RTMS92}.

\begin{figure}
\epsfxsize=\hsize
\centerline{\epsfbox{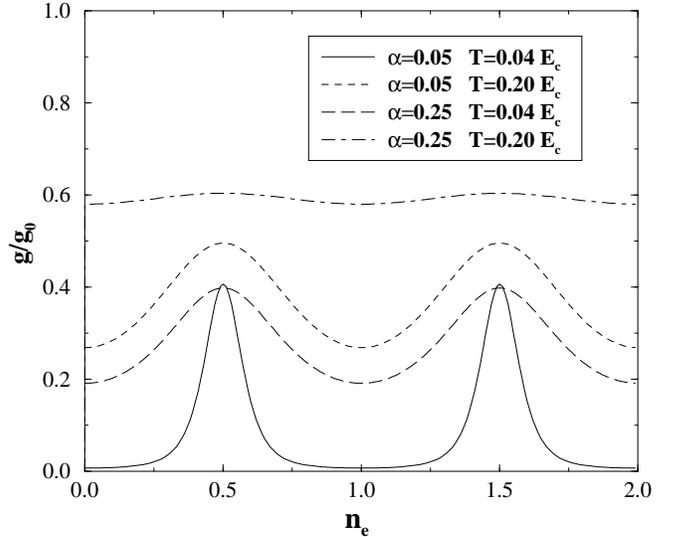}} %\vspace{-1.5cm}
\caption{Conductance of the single-electron transistor ($\epsilon=0$),
as a function of the dimensionless bias charge $n_e$, for several  
values of the coupling $\alpha$, and two different temperatures.}
\label{fig:conductance}
\end{figure}

We show the adequacy of the approximation,
eq.~(\ref{inter}), by plotting the conductances
estimated in this way, as a function of the bias charge $n_e$,
in fig.~(\ref{fig:conductance}).
\begin{figure}
\epsfxsize=\hsize
\centerline{\epsfbox{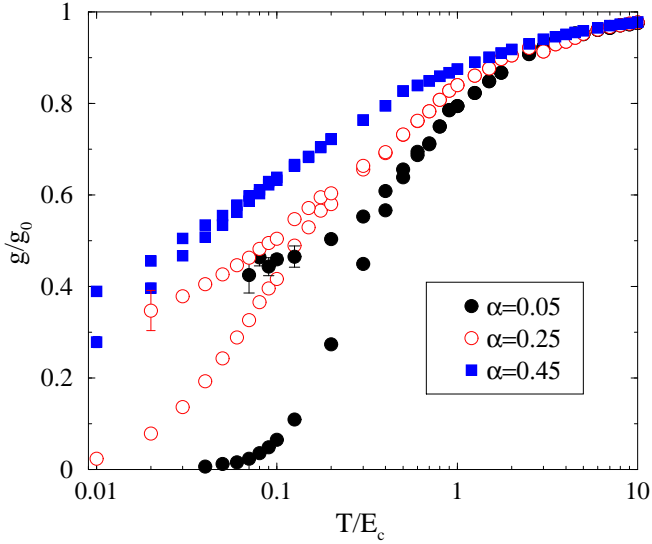}} %\vspace{-1.5cm}
\caption{Maximum and minimum values of
the conductance of the single-electron transistor ($\epsilon=0$), as
a function of temperature, for different values of the 
coupling $\alpha$.} 
\label{fig:maxmine0}
\end{figure}
The minimum ($n_e = 0$) and maximum conductances ($n_e = 1/2$)
for different values of $\alpha$ and temperatures are shown
in fig.~(\ref{fig:maxmine0}).                                
\subsection{$\epsilon \ne 0$}
We now extend the previous approximation to the conductance,
eq.~(\ref{inter}) to the case $\epsilon \ne 0$. The main modification
in eq.~(\ref{conductance}) is that a factor $\omega / ( 1 - e^{-\beta
\omega} )$ within the integral has to be replaced by the
effective tunneling density of states, $D_{\rm eff}$, given, at zero
temperature, by eq.~(\ref{effdosnew}). At finite temperatures,
the corresponding expression is approximately
\begin{equation}
D_{\rm eff} ( \omega ) \propto \hbox{max} [ T ( T / E_C )^{- \epsilon} , \omega  
( \omega / E_C )^{- \epsilon} ] \; .
\end{equation}
 The relevant range in the integrand in
eq.~(\ref{conductance}) is from $\omega = 0$ to $\omega \approx T$.
In the following, we will factor the $\epsilon$ dependent
part of the effective density of states, and we write the generalization
of eq.~(\ref{effdosnew}) to finite temperatures as:
\begin{equation}
D_{\rm eff} ( \omega ) = \frac{\omega}{1 - e^{-\beta \omega}}
D_{res} ( \epsilon , \omega )
\end{equation} 
Finally, when inserting this expression into eq.~(\ref{conductance}),
we make the approximation:
\begin{equation}
D_{res} ( \epsilon , \omega ) \approx \left( \frac{T}{E_C} 
\right)^{-\epsilon}
\end{equation}
With this approximation, we can perform the same analysis in the high and
low temperature regimes as before, to obtain the interpolation
formula:
\begin{equation}
g ( T ) \approx g_0 \left( \frac{T}{E_C} \right)^{-\epsilon}
G \left( \frac{\beta}{2} \right)
\end{equation}
This expression includes again the relevant physical processes at high
and low temperatures.

\begin{figure}
\epsfxsize=\hsize
\centerline{\epsfbox{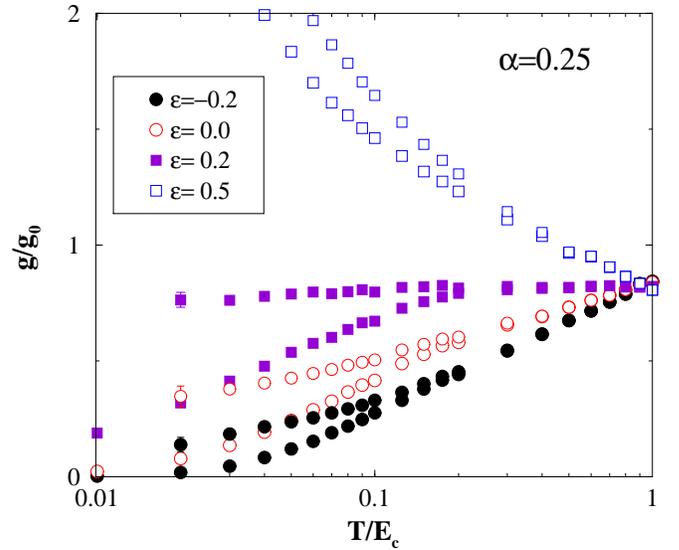}} %\vspace{-1.5cm}
\caption{Maximum and minimum values of
the conductance of the single-electron transistor,
for different values of $\epsilon$.}
\label{fig:exciton}
\end{figure}

Results for the maximum and minimum values of the conductances,
for different values of $\epsilon$, are presented in fig.~(\ref{fig:exciton}).
In the non phase-coherent regime, at very low temperatures, $T \ll E^*_C$, the conductance away  
from resonance should vary as $g \propto T^{2 - 2 \epsilon}$.
Exactly at resonance, $n_e = 1/2$, the conductance
diverges as $T^{-\epsilon}$.                                          
The most interesting result is the divergence of the conductance,
at low temperatures, for $\epsilon = 0.5$, where the excitonic effects
are strong enough to drive the system to the phase-coherent phase.

\begin{figure}
\epsfxsize=\hsize
\centerline{\epsfbox{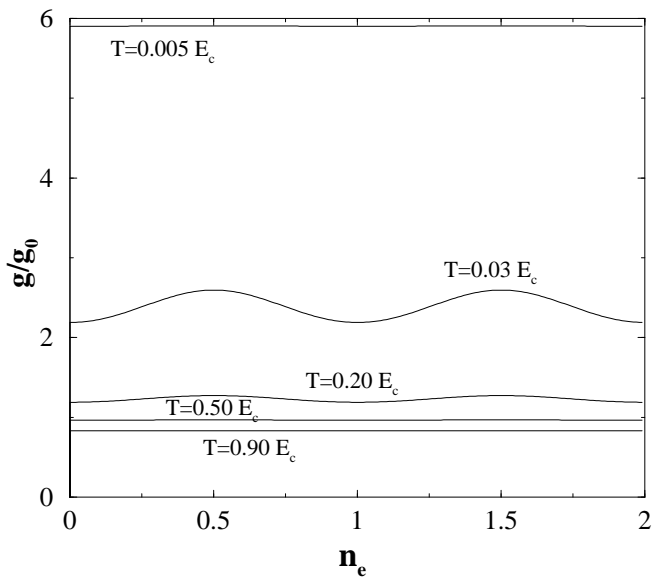}} %\vspace{-1.5cm}
\caption{Conductance, as a function of $n_e$ and temperature,
of a single-electron transistor with $\alpha = 0.25$ and $\epsilon = 0.5$.}
\label{fig:exciton1}
\end{figure}

\begin{figure}
\epsfxsize=\hsize
%\centerline{\epsfbox{exciton2.eps}} %\vspace{-1.5cm}
\centerline{\epsfbox{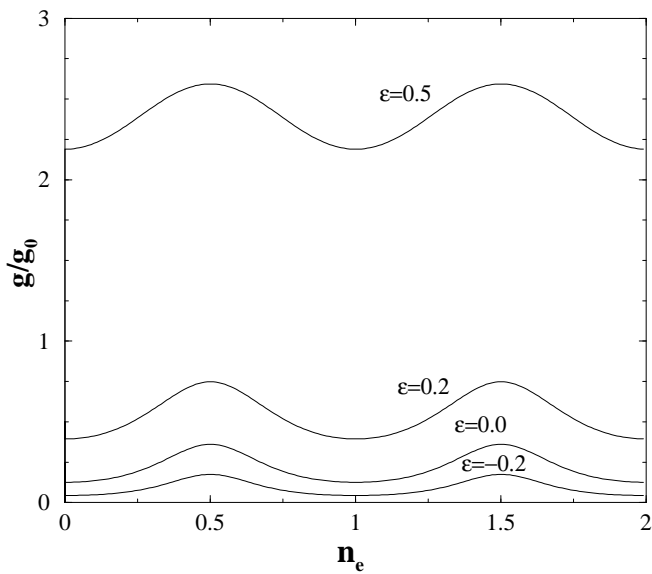}}   
\caption{Conductance, as a function of $n_e$ ,
of a single-electron transistor with $\alpha = 0.25$ and different
values of $\epsilon$ ($T = 0.03 E_C$).}
\label{fig:exciton2}
\end{figure}

The full conductance, as a function of $n_e$, is shown in 
fig.~(\ref{fig:exciton1}), for $\alpha = 0.25$ and $\epsilon = 0.5$.
As mentioned above, for these parameters the system is
already in the phase coherent regime.
The conductance behaves in a way similar to that in the usual case
($\epsilon = 0$), and a peaked structure develops.
The absolute magnitude, however, increases as
the temperature is lowered. Note that the effective charging energy
is finite as $T \rightarrow 0$ (see fig.~(\ref{fig:eca25})).
It is interesting to note that, in this phase with complete
suppression of Coulomb blockade effects at low temperatures
(high values of $\epsilon$ and high conductances), the
peak structure appears only for an intermediate range
of temperatures, and it is washed out at very low temperatures.
In fig.~(\ref{fig:exciton2}) we show the conductance as
a function of $\epsilon$ for a fixed temperature. It is evident the increase of the 
conductance as $\epsilon$ increases.                                                 

\section{Discussion.}  
In the following, we discuss
some experimental evidence which can be explained within 
the model discussed here.  

It has been pointed out\cite{BG99} that the correlations between
the conductances for neighboring charge states of a quantum dot\cite{Cetal97}
are too weak to be explained using standard methods for
disordered, non-interacting systems. The experiments reported
in\cite{Cetal97} are in the cotunneling regime. In the
presence of non-equilibrium processes, we expect a behavior
of the type $g \propto T^{2 - 2 \epsilon}$. 
Note that $\epsilon$ is determined by phase shifts, which depend on
microscopic details of the contacts. Thus, it can be expected to vary with
the charge state of the dot, and to lead to large differences in
the conductances of neighboring valleys. 

It has been shown that the temperature dependence of the conductance quantum dots,
away from resonances,
can be opposite to that expected in a system exhibiting
Coulomb blockade\cite{Metal99}. This effect could not be
attributed to Kondo physics, as the data do not show an even-odd
alternation. The reported behavior can be explained within our
model, assuming that the value of $\epsilon$ is sufficiently large,
and dependent on the charge state of the dot.

In ref.\cite{Fetal98} the inelastic contribution to the conductance 
in a double dot sytem, where the electronic states in the two
dots are separated by an energy $\epsilon$ is measured.
The result is approximately given by
$I_{\rm inel} ( \epsilon ) \propto \epsilon^{\lambda}$, 
where $\lambda$ is a negative constant of
order unity. 
Taking into account only  one
electronic state within each dot, the problem can be reduced
to that of a dissipative, biased two level system. 
The inelastic conductance reflects the nature of the low
energy excitations coupled to the two level system. The observed power
law decay implies that the spectral strength of the coupling, that is
the function $J ( \omega )$ in the standard literature\cite{W93}, should
be ohmic, $J ( \omega ) \propto | \omega |$.
This has led to the proposal that the excitations coupled to
the charges in the double dot system are piezoelectric phonons\cite{Fetal98,BK99}.
It is interesting to note that the excitation of electron-hole
pairs leads also to an ohmic spectral function.
Thus, at sufficiently low energies,
an orthogonality catastrophe due to electron-hole pairs
shows a behavior
indistinguishable from that arising from piezoelectric phonons.
The contributions from the two types of excitations can be distinguished
at the natural cutoff scale for phonons, which
is the energy of a phonon whose wavelength is of the order of the 
dimensions of the device.  On the other hand, the simplest 
prediction for the expected behavior of the current induced by
the emission of the electron-hole pairs is:
\begin{equation}
I ( \epsilon ) 
\propto K \frac{1}{1 - e^{- \beta \sqrt{\Delta_0^2 + \epsilon_b^2}}}
\frac{\Delta_0^2}{\sqrt{\Delta_0^2 + \epsilon_b^2}}
\label{rate}
\end{equation}
where $\Delta_0$ is the tunneling element between the two dots,
$\epsilon_b$ is the bias, and $K$ is the coupling constant
(referred to as $\epsilon$ in other sections of this paper).
This expression gives the absorption rate of a dissipative two
level system in the weak coupling regime, $K \ll 1$\cite{W93}.
The natural cutoff for electron-hole pairs is bounded by the charging energy
of the system.
It would be interesting to disentangle the relative contributions
of electron-hole pairs and piezoelectric phonons to the inelastic
current.

The photo-induced conductance in a double dot system has also been
measured\cite{Betal98b}. The analysis of the contribution of inelastic processes
due to electron hole pairs
to the measured conductance proceeds in the same way as in the interpretation
of the previous experiment. Let us suppose that a photon of
energy $\omega_{ph}$ excites
an electron within one dot. Assuming that
the coupling to the environment is weak,
the rate at which the electron tunnels to the second dot
by losing an energy $\epsilon$ is given by
eq.(\ref{rate}). The experiments in\cite{Betal98b}
suggest that the number of states within each dot are discrete.
Then, the induced conductance should show a series of peaks, related
to resonant photon absorption within one dot. The height of each peak
is determined by the decay rate to lower excited states in the other dot,
and it can be written as a sum of terms with the dependence
given in eq.(\ref{rate}), where $\epsilon$
is the energy difference between the initial and final states.
The envelope of
the spectrum should look like a power law, in qualitative agreement
with the experiments.

It is interesting to note
that bunching of energy levels in quantum dots have been reported\cite{ZAPW97}.
The separation between peaks defines the charging energy, which, according
to the experiments, vanishes for certain charge states. This behavior
can be explained if the excitonic effects drive the quantum dot
beyond the transition, and charging effects are totally suppressed.    
This mechanism can also play some role in the observed
transitions in granular wires\cite{DRG98,Hetal96}.
\section{Conclusions.}
We have analyzed the effects of non-equilibrium transients
after a tunneling process on the conductance of quantum dots.
They are related to the change in the electrostatic potential 
of the dot upon the addition of a single electron. These effects can enhance or suppress the Coulomb blockade.
The  most striking effect arise
from the formation of an exciton-like resonance at the
Fermi level after the charging process and lead to the complete suppression of the Coulomb blockade and a diverging conductance
at low temperatures. It appears for sufficiently large values of the conductance, $\alpha$ , and the non-equilibrium phase shifts which define $\epsilon$ in our model. The same potential
which leads to this dynamic resonance plays a role in the
deviations of the level spacings from the standard Coulomb
blockade model\cite{BMM97}.

For simplicity, we have
considered the simplest case, a single-electron transistor.
The analysis reported here can be extended, in a straightforward
fashion to other devices, like double quantum dots,
where the effects described
here should be easier to observe.

As discussed, the non-equilibrium effects considered in this paper may have been
 observed already in the conductance of quantum dots. 

\section{Acknowledgments}
One of us (E. B.) is thankful to the University of Karlsruhe for
hospitality. We acknowledge financial support from CICyT (Spain)
through grant PB0875/96, CAM (Madrid) through grant 07N/0045/1998 and FPI,
and the European Union through grant ERBFMRXCT960042.

\end{document}